\documentclass{PoS}

\newcommand{ \be}{\begin{equation}}
\newcommand{ \ee}{\end{equation}}
\newcommand{ \bea}{\begin{eqnarray}}
\newcommand{ \eea}{\end{eqnarray}}
\newcommand{ \mysmall}[1]{\scriptscriptstyle #1} 

\newcommand{ \mw}{M_{\mysmall{W}}}

\newcommand{ \ct} {c_{\mysmall{W}}}
\newcommand{ \st} {s_{\mysmall{W}}}
\newcommand{ \stbar} {\bar{s}_{\mysmall{W}}}

\def\chic#1{{\scriptscriptstyle #1}}

\title{Neutrino-Nuclear Coherent Scattering \\ 
and the Effective Neutrino Charge Radius}

\ShortTitle{Neutrino-Nuclear Coherent Scattering and the Effective Neutrino
Charge Radius}

\author{\speaker{Joannis Papavassiliou}\\
         Departamento de F\'\i sica Te\`orica and IFIC Centro Mixto,
        Universidad de Valencia--CSIC, E-46100, Burjassot, Valencia, 
	Spain\\
        E-mail: \email{Joannis.Papavassiliou@uv.es}}

\author{Jose Bernab\'eu\\
         Departamento de F\'\i sica Te\`orica and IFIC Centro Mixto,
        Universidad de Valencia--CSIC, E-46100, Burjassot, Valencia, 
	Spain\\
        E-mail: \email{Jose.Bernabeu@uv.es}}

\author{Massimo Passera\\
         Dipartimento di Fisica ``G.~Galilei'', Universit\`{a} 
        di Padova and  INFN, Sezione di Padova, I-35131, Padova, Italy\\
         E-mail: \email{massimo.passera@pd.infn.it}}

\abstract{We propose to extract the value of the effective neutrino charge
radius from the coherent scattering of a neutrino against a heavy nucleus.
In such an experiment the relevant quantity to measure is the kinetic energy
distribution of the recoiling nucleus, which, in turn, may be directly
related to the shift in the value of the effective weak mixing angle
produced by the neutrino charge radius.  This type of experiment has been
proposed in order to observe the coherent elastic neutrino-nuclear
scattering for the first time.  If interpreted in the way suggested in this
work, such an experiment would constitute the first terrestrial attempt to
measure this intrinsic electromagnetic property of the neutrino.}

\FullConference{International Europhysics Conference on High Energy Physics\\
		 July 21st - 27th 2005\\
		 Lisboa, Portugal}

\begin{document}

It is well-established by now that the difficulties associated with the
definition \cite{Lucio:1984mg} of the neutrino charge radius ({\small NCR})
have been conclusively settled in a series of papers
\cite{Bernabeu:2000hf,Bernabeu:2002nw,Bernabeu:2004jr}, by resorting to the
well-defined electroweak gauge-invariant separation of physical amplitudes
into effective self-energy, vertex and box sub-amplitudes, implemented by
the pinch technique formalism \cite{Cornwall:1982zr}.  Thus, within the
Standard Model, at one-loop order, the {\small NCR}, to be denoted by $\big
<r^2_{\nu_i}\, \big>$, is given by
\be
\big <r^2_{\nu_i}\,  \big> =\, 
\frac{G_{\chic F}}{4\, {\sqrt 2 }\, \pi^2} 
\Bigg[3 
- 2\log \Bigg(\frac{m_{\chic i}^2}{M_{\chic W}^2} \Bigg) \Bigg]\, ,
\label{ncr}
\ee
where $i= e,\mu,\tau$, the $m_i$ denotes the mass of the charged iso-doublet
partner of the neutrino under consideration, and $G_{\chic F}$ is the Fermi
constant.  In addition, as has been demonstrated in \cite{Bernabeu:2002nw},
the {\small NCR} so defined can be expressed in terms of a judicious
combination of physical cross-sections, a fact which promotes it into a
genuine physical observable. This possibility has revived the interest in
this quantity \cite{PDB}, and makes the issue of its actual experimental
measurement all the more interesting.  In this talk we will argue that
upcoming experiments involving coherent neutrino-nuclear scattering
\cite{Giomataris:2005bb,Barranco:2005yy} may provide the first opportunity
for measuring the {\small NCR} (or, at least, for placing bounds on its
value).

The notion of coherent nuclear scattering 
is well-known from electron scattering. In the neutrino case it was developed 
in connection with the discovery of weak neutral currents, with a component 
proportional to the number operator \cite{Bernabeu:1975tw}. When a
projectile (e.g.\ a neutrino) scatters elastically from a composite system
(e.g.\ a nucleus), the amplitude $F({\bf p'},{\bf p})$ for scattering from
an incoming momentum ${\bf p}$ to an outgoing momentum ${\bf p'}$ is given 
as the sum of the contributions from
each constituent,
\be
F({\bf p'},{\bf p})= \sum_{j=1}^{A} f_j({\bf p'},{\bf p}) 
e^{i {\bf q}\cdot {\bf x}_j} \,\, ,
\label{coh1}
\ee
where ${\bf q}={\bf p'}-{\bf p}$ is the momentum transfer and the individual
amplitudes $f_j({\bf p'},{\bf p})$ are added with a relative phase-factor, 
determined by the corresponding wave function.
The differential cross-section is then
\be 
\frac{d\sigma}{d\Omega} = |F({\bf
p'},{\bf p})|^2 = \sum_{j=1}^{A} |f_j({\bf p'},{\bf p})|^2 +
\sum_{j,i}^{i\neq j} f_i({\bf p'},{\bf p})f_j^{\dagger}({\bf p'},{\bf p})
e^{i {\bf q}\cdot ({\bf x}_j-{\bf x}_i)} \, .
\label{coh2}
\ee
In principle, due to the presence of the phase factors, major cancellations
may take place among the $A(A-1)$ terms in the second (non-diagonal) sum.
This happens for  $qR \gg 1$, where  $R$ is the size of the composite system,  
and the scattering would be incoherent.
On the contrary, under the condition that $qR \ll 1$, 
then all phase factors may be approximated by unity, and the terms
in (\ref{coh2}) add coherently.  If there were only one type of constituent,
i.e.\ $f_j({\bf p'},{\bf p}) = f({\bf p'},{\bf p})$ for all $j$, then
(\ref{coh2}) would reduce to
\be
\frac{d\sigma}{d\Omega} = A^2 \left|f({\bf p'},{\bf p})\right|^2 
\ee
Evidently, in that case, the {\it coherent} scattering cross-section would
be enhanced by a factor of $A^2$ compared to that of a single constituent.
In the realistic case of a nucleus with $Z$ protons and $N$ neutrons,
and assuming zero nuclear spin, the corresponding
differential cross-section reads~\cite{Bernabeu:1975tw}
\be
\frac{d\sigma}{d\Omega} = \frac{G^2_{{\mysmall F}}}{4(2\pi)^2} E^2 
(1+\cos\theta) \left[(1-4 \st^2)Z -  N\right]^2,
\label{dsdy}
\ee
where $\st$ is the sine of the weak mixing angle, $d\Omega = d\phi\,
d(\cos\theta)$, and $\theta$ is the scattering angle.  
For elastic scattering, the scattering angle is related to the nuclear recoil, 
so that 
the kinetic energy distribution of the recoiling nucleus is written as
\cite{Bernabeu:1975tw}
\be
\frac{d\sigma}{dy} = \frac{G^2_{{\mysmall F}}}{2\pi}
\frac{M(M+2E)^2}{[M+2E(1-y)]^3} E^2 (1-y) \left[(1-4 \st^2)Z -N\right]^2,
\label{dsdyapp}
\ee
where $M$ is the mass of the nucleus, $y= T/T_{{\mysmall max}}$, $y \in
[0,1]$ , and $T_{{\mysmall max}} = {2 E^2 }/{ (M + 2 E)}$.  For $2 E \ll M$,
$T_{{\mysmall max}} \simeq 2 E^2 / M$, and, to an excellent approximation,
(\ref{dsdy}) simplifies to the rather compact expression,
\be
\frac{d\sigma}{dy} \simeq \frac{G^2_{{\mysmall F}}}{2\pi}
E^2 (1-y) \left[(1-4 \st^2)Z -  N\right]^2.
\label{dsdysim}
\ee
%


\begin{figure}[]
\begin{center}
\includegraphics[width=10cm]{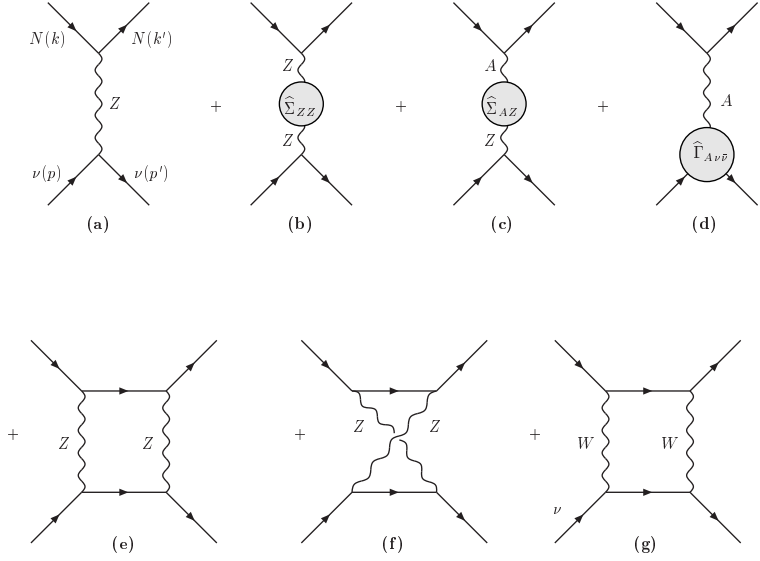}
\caption{\label{Fig1}   
The various diagrams contributing to the one-loop amplitude.
One-loop vertex corrections with an incoming $Z$ are omitted, 
because they are suppressed by a factor $q^2/M_{Z}^2$ ~\cite{Bernabeu:2002nw}.
}
\end{center}
\end{figure}

The one-loop interactions between a nucleon $N$ and a neutrino $\nu$ are
shown in Fig.1; of course, when the nucleon is a neutron, both ({\bf c})
and ({\bf d}) vanish. 
It is important to realize that, in the kinematic
limit considered (i.e. $|q^2| \ll M^2$), all one-loop contributions may be
absorbed into shifts of the original parameters appearing in the Born
amplitude (\ref{dsdysim}), giving rise to a Born-improved amplitude.  In
particular, as has been explained in detail in \cite{Bernabeu:2002nw},
diagrams ({\bf a}), ({\bf b}), and ({\bf c}), combine to form two
renormalization-group-invariant quantities,
\be
\bar{R}_{\chic{Z}}(q^2) = 
\frac {\alpha_{\mysmall{W}}}{\ct^2}
\bigg[q^2 - M_\chic{Z}^2 +\Re e\,\{\widehat{\Sigma}_{\chic{Z}\chic{Z}}(q^2)\}
\bigg]^{-1} ,\,\,\,\,
\stbar^2(q^2) =  \st^{2}\Biggl(1 - \frac{\ct}{\st}\, 
\Re e\,\{\widehat{\Pi}_{\chic{A}\chic{Z}}(q^2)\}\Biggr) \,,
\label{RW}
\ee
where $\alpha_{\mysmall{W}} = g_{\mysmall{W}}^2/4\pi$, $\Re e\,\{...\}$
denotes the real part, and $\widehat{\Sigma}_{\chic{A}\chic{Z}}(q^2) = q^2
\widehat{\Pi}_{\chic{A}\chic{Z}}(q^2)$.  $\bar{R}_{\chic{Z}}(q^2)$ can be
directly related to the effective electroweak (running) coupling, and
eventually be interpreted as a shift to $G^2_{\mysmall F}$. Similarly,
$\stbar^{2}(q^2)$ defines the effective (running) weak mixing angle. It
turns out that the UV-finite contribution from the {\small NCR}, contained
in diagram ({\bf d}), may be also absorbed into a shift of $\st^{2}$. In
fact, a detailed analysis based on the methodology developed in
\cite{Hagiwara:1994pw}, reveals that, in the kinematic range of interest,
the numerical impact of $\bar{R}_{\chic{Z}}(q^2)$ and $\stbar^{2}(q^2)$ is
negligible, i.e.\ these quantities do not run appreciably.  Instead, the
contribution from the {\small NCR} amounts to a correction of few percents
to $\st^{2}$, given by an expression of the form
$\st^{2} \longrightarrow \st^{2} 
\bigg(1-\frac{2}{3}\, \mw^{2} \, \big <r^2_{\nu_i}\,  \big>\bigg)$.
Finally, the contributions of the boxes are to be included.  One can show
that the sum of ({\bf e}) and ({\bf f}) vanishes in the relevant kinematic
limit, whereas graph ({\bf g}) gives a contribution proportional to
$g_{\mysmall{W}}^4/\mw^{2}$, whose impact is currently under investigation.
Finally we would like to point out that if one were to consider the
differences in the cross-sections between two different neutrino species
scattering coherently off the same nucleus, as proposed by Sehgal two
decades ago \cite{Sehgal:1985iu}, one would eliminate all unwanted
contributions, such as boxes, thus measuring the {\it difference} between
the two corresponding charge radii. Such a difference would also contribute 
to a difference for the neutrino index of refraction in nuclear matter
\cite{Botella:1986wy}.

\subsection*{Acknowledgments}
This work was supported by the MCyT grant FPA2002-00612 and by the European
Program MRTN-CT-2004-503369.

\end{document}